\documentclass{iopconfser}

\usepackage{amssymb,amsfonts}

\newcommand{\be}{\begin{equation}}
\newcommand{\ee}{\end{equation}}
\newcommand{\bea}{\begin{eqnarray}}
\newcommand{\eea}{\end{eqnarray}}
\newcommand{\nn}{\nonumber}

\begin{document}

\title{Asymptotics and Universality in Black Holes: 
    from the quasinormal Weyl's law to the binary merger
    waveform}

\author{J.L Jaramillo$^{1}$, L. Al Sheikh$^{2,1}$, J. Besson$^{1,3,4}$,
  B. Krishnan$^{5,3}$, M. Lenzi$^{6,7}$, R.P. Macedo$^{8,9}$,
  O. Meneses-Rojas$^{1}$, B. Raffaelli$^{1}$, C.F. Sopuerta$^{5,6}$, C. Vitel$^{1}$}

\affil{$^1$Institut de Math\'ematiques de Bourgogne UMR 5584,
  Universit\'e Bourgogne Europe, CNRS, F-21000 Dijon, France}
\bigskip
\affil{$^2$J\"ulich Centre Neutron Science (JCNS-4), Forschungszentrum
  J\"ulich, Garching, Germany}
  \bigskip
\affil{$^3$Albert-Einstein-Institut, Max-Planck-Institut für Gravitationsphysik, Callinstraße 38, 30167 Hannover, Germany}
\bigskip
\affil{$^4$Leibniz Universit\"at Hannover, 30167 Hannover, Germany}
\bigskip
\affil{$^5$Institute for Mathematics, Astrophysics and Particle Physics,
Radboud University, Heyendaalseweg 135, 6525 AJ Nijmegen, The Netherlands}
\bigskip
\affil{$^6$Institut de Ci\`encies de l’Espai (ICE, CSIC), Campus UAB,
Carrer de Can Magrans s/n, 08193 Cerdanyola del Vall\`es, Spain}
\bigskip
\affil{$^7$Institut d’Estudis Espacials de Catalunya (IEEC), Carrer Esteve Terradas,
1, Edifici RDIT, Campus PMT-UPC, 08860 Castelldefels, Spain}
\bigskip
\affil{$^8$Niels Bohr International Academy, Niels Bohr Institute, Blegdamsvej 17, 2100 Copenhagen, Denmark}
\bigskip
\affil{$^9$School of Mathematical Sciences and STAG Research Centre, University of Southampton, Southampton, SO17 1BJ, United Kingdom}

\medskip

\email{Jose-Luis.Jaramillo-Martin@ube.fr}

\medskip


\begin{abstract}
  Current state-of-the-art approaches to black hole (BH) dynamics,
  encompassing
  several effective approximation schemes,
  offer a remarkable control of the quantitative aspects of strong gravity.
  They also provide key insights into some qualitative aspects of the problem. In spite of this, there remain blind spots that hinder the understanding of the mechanisms underlying some observed phenomena, in particular concerning simplicity and universality in BH spacetimes. Adopting an `asymptotic reasoning' approach, 
  by filtering non-essential degrees of freedom, can potentially unveil universality patterns by identifying key underlying structural stability mechanisms. We first illustrate such an asymptotic approach by focusing on
  a BH quasinormal (QNM) Weyl's law, that accounts for the universal asymptotics of the QNM ``counting function''.
  This permits to identify light-trapping and the (local) redshift effect as the underlying mechanisms, also offering a bridge to the universal patterns found in BH QNM spectral instability. 
  As a by-product, Weyl's law universality formally opens an observational access to spacetime (effective) dimensionality. More heuristically, we sketch a program recently put forward
  to apply such  `asymptotic reasoning' to address the observed simplicity and universality patterns in binary BH merger dynamics. This program is built as a hierarchy of asymptotic models, potentially making contact with integrability theory in gravity, namely through the background sector in a ``wave-mean flow'' approach to BH
  binary dynamics.
\end{abstract}


\section{Black Hole simplicity and universality: `asymptotic reasoning' and `structural stability'}
Our motivational guiding problem is the stark contrast between early
educated guesses \cite{Schutz:2004uj}
for the binary black hole (BBH) gravitational waveform and the actual BBH waveform  eventually obtained
by numerical relativity \cite{Pre05} and then
observed by interferometric antennae \cite{Abbott:2016blz}: whereas the former shows complicated patterns
in the merger phase, the latter is remarkable by its simplicity.
A natural question is posed:

\medskip

\centerline{\em Is the binary black hole merger waveform elegant or, rather, boring?}

\medskip

\noindent The BBH waveform is not only simple, but is also universal (in the quasi-circular class). 
We address such simplicity and universality in terms of the notions of
{\em asymptotic reasoning}~\cite{batterman2002devil,Batte97} and {\em structural stability}:
\begin{itemize}
\item[i)] {\em Simplicity: asymptotic reasoning.} Instead of attempting a `full and
  exact' quantitative description, we filter the overload of details encumbering
  structural features, aiming at the underlying qualitative mechanisms.
  This step involves a (simplifying) `zoom' into a  small-parameter  asymptotic regime.

\item[ii)] {\em Universality: structural stability}. 
  The focus on structurally stable
  (stable properties of generic configurations)
  features of the underlying mechanisms, once dwelling in simplified asymptotic regimes, typically
  permits to unveil {\em universal patterns}  
  persisting in the theory beyond the asymptotic regime.
  
\end{itemize}
We can now formulate our ultimate goal, that provides the leitmotiv for the research described here: \\
To identify the mechanisms underlying the observed `simplicity and universality' in BBH mergers
and assess their role as `keys/probes' into the fundamental theoretical structures
ruling gravitational dynamics.

\section{The Black Hole Quasinormal Weyl's law: a case of asymptotic reasoning}
Before we proceed to give more details on the BBH program (admittedly a bold proposal),
we illustrate the sketched `asymptotic reasoning' methodology in a   sober example,
involving the counting of black hole (BH) quasi-normal mode (QNM) complex frequencies $\omega_n$,
where sound results can be presented~\cite{Jaramillo:2021tmt,Jaramillo:2022zvf,Besson:2024adi}.

\subsection{The classical Weyl' law: an instance of the `asymptotic reasoning' path to universality}
We start by reviewing the classical Weyl's law for the Laplacian operator.
Given a compact $d$-dimensional domain $D$, namely a compact cavity,
we consider the spectral problem  $-\Delta\phi_n = \lambda_n\phi_n$ (with `reflecting' boundaries), and
 define the eigenvalue counting function
$N(\lambda)$ as
$N(\lambda) = \#\{ \lambda_n\in\mathbb{R}, \hbox{ such that } \lambda_n\leq \lambda\}$.

 Obtaining the `full and exact' expression for $N(\lambda)$ is indeed a most challenging problem.
 However, in the large-$\lambda$ asymptotic limit (or the small-$(1/\lambda)$ limit)
 the situation drastically simplifies and a closed (asymptotic) form is actually obtained. The latter
 is provided by the Weyl's law~\cite{Weyl11,Weyl12} (cf. \cite{Kac66,Arendt2009,Ivrii16}) 
\bea
\label{e:counting_function_compact}
N(\lambda) \sim C_d \mathrm{Vol}_d(D) \lambda^{\frac{d}{2}} +  o(\lambda^{\frac{d-1}{2}}) \ \ , \ \ (\lambda\to\infty) \ ,
\eea
with constant $C_d=\mathrm{Vol}_d(B^d_1)/(2\pi)^d$, where $\mathrm{Vol}_d(B^d_1)$ is the
Euclidean volume of a $d$-dimensional ball
of unit radius. The expression (\ref{e:counting_function_compact}) is not only {\em simple}, it is
actually also {\em universal}, only depending on the volume of the domain $D$ and its dimensionality.
Indeed, such universality played a key role in establishing the universality of the black body radiation
\cite{Kac66}. Weyl's law provides an instance of asymptotic reasoning.

\subsection{The Schwarzschild BH case: insights from a `poor's man' but enlightening approach}
In the spirit of the previous discussion, we explore \cite{Jaramillo:2021tmt,Jaramillo:2022zvf}
a Weyl-like law for BH QNMs  analogous to
(\ref{e:counting_function_compact})
\bea
\label{e:BH_QNM_Weyl_v2}
N(\omega) \sim \frac{2}{c^d}C_d\mathrm{Vol}_d^{\mathrm{eff}} \omega^d +  o(\omega^{d-1})
\ \ , \ \ (\omega\to \infty, \ \ \hbox{with } \lambda=\omega^2) \ ,
\eea
where $\mathrm{Vol}_d^{\mathrm{eff}}$ is now an effective volume.
In particular, we aim
at assessing if such a BH QNM Weyl's law can provide a possible probe into some structural universal aspects in the scattering on BH spacetimes.

We first consider the ($d=3$) Schwarzschild BH toy-model, 
making key use of the asymptotic expressions \cite{Nollert:1993zz,Kokkotas:1999bd}
(with $\kappa$ the BH surface gravity, $\kappa=1/(4M)$ in Schwarzschild;
$G=c=1$ here throughout)
\bea
\omega_{n\ell m} \sim \pm \frac{\ln(3)}{8M\pi} +
                  i\kappa\left(n - \frac{1}{2}\right) + \ldots \  (n \to \infty) \ \ , \ \
 \mathrm{Re}(\omega_{0,\ell}) \sim \frac{1}{3\sqrt{3}M} \left(\ell + \frac{1}{2}\right) + \ldots  \ (\ell\to\infty) \ .
\eea
The crucial point is that the first expression
is controlled by the surface gravity $\kappa$, whereas the existence of a ``light ring''
(or ``photon sphere'', with area $\mathrm{Area}^{\mathrm{LR}}$)
is the structure underlying the second one.
Following an asymptotic reasoning argument (cf. details in \cite{Jaramillo:2022zvf}),
the $N(\omega)$ asymptotics is expressed as  
       \bea
\label{e:BH_QNM_Weyl_v4}
N(\omega) &\sim&
\frac{1}{c^2} \left(\frac{1}{6\pi^2}\right)  \left(\left(\frac{8\pi}{\kappa}\right) \cdot
\mathrm{Area}^{\mathrm{LR}}\right) \omega^3 +  o(\omega^{2}) \ .
\eea
In sum, such asymptotic reasoning unveils an expression for $N(\omega)$
in terms of a power-law in $\omega$ with a coefficient factorised into ``radial'' and ``angular'' parts, with 
the following structural features:
 i) the ``radial'' part involves a  ``thermalization'' time characterised by the
 surface gravity $\kappa$, ii) the ``angular'' part is controlled by the
 light-ring (``trapped set'') area, iii) the power is fixed by the space(time) dimension. 

              
 \subsection{The generic case: (local) redshift effect and light trapping as underlying mechanisms}
 In spite of the ``poor's man'' flavour of the previous  discussion,
 such asymptotic reasoning has the virtue of having identified
 some underlying structures that do generalise to the generic case. Specifically:
 i) the (stable) half-plane of BH QNM complex frequencies is structured in horizontal bands
 of width $\kappa$ \cite{Warnick:2013hba}, ultimately due to the so-called (local) redshift effect
 \cite{Dafermos:2005eh,Dafermos:2008en},
 ii) the counting of the BH QNM in a given band (namely, the fundamental one)
 is determined by the (phase space) volume of trapped region  \cite{Dyatlov:2013hba,Dyatlov:2013fua}. This permits
 to reprise the reasoning in the Schwarzschild case, leading to the following
 conjecture~\cite{Jaramillo:2022zvf}:

  
 

  {\em The asymptotics of the
  QNM counting function for a  generic ($d+1$)-dimensional stationary BH is}
\bea
\label{e:BH_QNM_Weyl_dim-d}
N(\omega) \sim \frac{1}{(2\pi)^d}
\left(\left(\frac{8\pi}{\kappa}\right) \cdot \mathrm{Vol}_{K_t}(\xi_t^2\leq 1)\right) \omega^d  +
o(\omega^{d-1}) \ , \nn
\eea
{\em where $\kappa$ is the BH surface gravity and $K_t$ is the $(d-1)$-dimensional symplectic manifold
  determined by constant time sections of the trapped set $K$ in phase space
  (with canonical-pair coordinates $(x^a, \xi_a)$).}

The `simplicity and universality' of the proposed BH QNM Weyl's law is explained
in terms of the following qualitative, structurally-stable mechanisms: 
 i) the {\em local redshift effect} controlled by the surface gravity $\kappa$,
 ii) the {\em light trapping} (in phase space), and iii) the role of {\em dimensionality}.
Such BH QNM Weyl's law offers an instance of the asymptotic reasoning methodology
 unveiling the key underlying mechanisms.


\section{A BBH merger waveform program: from asymptotics to integrability}
After illustrating asymptotic reasoning with the BH QNM Weyl's law case, we come back
to the simplicity and universality in the BBH merger waveform, with a focus on exploring the underlying mechanisms.

A key remark is the following: BBH mergers seem to behave more `linearly' than a priori expected.
We certainly do not claim that non-linear terms are not key for a full quantitative account (cf. e.g. recent
works on second-order QNMs), but we rather do argue  that adopting an ``effective linearity''
hypothesis proves fruitful to understand some underlying structures of 
the dominating BBH qualitative features\footnote{For a discussion of such 
``effective linearity'', see section I.B.3
of \cite{Jaramillo:2022mkh},  section 1 of \cite{Jaramillo:2022kuv} or section 8.2.2 of
\cite{JarLam24}, specially the `transparency' phenomenon 
in general relativity unveiled by Choquet-Bruhat~\cite{ChoBru69,Touati:2022cjx,Touati:2022btx}
and sketched in point (ix) of section 8.2.2 in \cite{JarLam24}. See also the
``the not-so-nonlinear nonlinearity of Einstein's equation'' discussed by Harte in \cite{Harte14}.}.

\subsection{BBH simplicity and universality: the `diffraction on caustics' mechanism}
The notion of `universal wave patterns' in dispersive partial differential equations (PDE)
\cite{Mille16} offers a tantalizing first hint to address the simplicity and universality
of the  BBH waveform
pattern.
This notion provides a PDE analogue of universality features in (critical) phase transitions
in statistical mechanics.

An archetype of such a phenomenon is furnished by the Burgers’ equation
modeling shocks in hydrodynamics.
Specifically, Burgers' equation regularized with a viscosity term proportional to an asymptotic
small-parameter $\epsilon$, writes
$\displaystyle \partial_t u^\epsilon + u^\epsilon\partial_x u^\epsilon = \epsilon \partial^2_{xx} u^\epsilon$.
By performing a `zoom' into the shock by making $\epsilon\to 0$, one concludes  
that all viscous shocks ``look the same'', described by the universal function
$\mathrm{tanh}(x)$ (cf. e.g. \cite{landau2013fluid,Jaramillo:2022oqn}),
independent of the initial data.
This offers a neat illustration of asymptotic reasoning.

In the same spirit, a universal behaviour occurs for arbitrary linear dispersive wave PDEs,
near `caustics'. Specifically, such caustics are regularised in a universal diffraction
pattern provided by the Airy function. For concreteness, one considers the semiclassical
(`WKB-like') treatment of the PDE with a small parameter $\epsilon$,  
$\psi(t,x) \sim \frac{e^{-i\pi/4}}{\sqrt{2\pi\epsilon t}}\int_{-\infty}^\infty e^{iI(k;x,t)/\epsilon}
\sqrt{\rho_0(k)} dk$. Characteristic lines (giving rise to caustics) are determined by
$I'(k;x,t)=0$. The generic caustic at $x_c$ is a `fold' with
$I(y;x_c,t)\sim \phi_{\mathrm{fold}}(\tau(t); k)$,
leading to the universal Airy diffraction pattern \cite{Jaramillo:2022mkh}
(the Fourier transform of the
dispersive $e^{ik^3/3}$)
\bea
  \frac{1}{2\pi}\int_{-\infty}^\infty dk \; e^{i \phi_{\mathrm{fold}}(\tau(t); k)}
  = \frac{1}{2\pi}\int_{-\infty}^\infty dk \; e^{i(\tau(t) k+k^3/3)} = \mathrm{Ai}(\tau(t)) \ .
  \eea

  {\bf Proposal of  Mechanism 1}. {\em Airy function as (linear) universal diffraction pattern on a `fold caustic',
  namely the special function in a `universal-wave-pattern' critical PDE phenomenon for  BBH waveforms.}

\subsection{BBH simplicity and universality: a hint into `integrability' as a fundamental mechanism}
The Airy function is a solution of the Airy equation, namely 
$\displaystyle \ddot{u}  - t u = 0$. The latter is the linearisation of
the so-called Painlev\'e-II equation, 
$\displaystyle \ddot{u}- t u \; -2 \; u^3  = 0$.
Its interest is that Painlev\'e transcendents (solutions to the six
Painlev\'e equations~\cite{Clark03,ConMus08}) are a smoking gun of the presence of an integrability structure.

Remarkably, the Painlev\'e-II equation shows up in BBH dynamics (see \cite{Jaramillo:2022oqn}):
i) in the inspiral phase (in the extreme-mass ratio case)
the Painlev\'e-II equation offers an analytical avenue to orbital motion  subject
to radiation-reaction~\cite{Rajeev:2008sw}, with its Painlev\'e-I contraction controlling the plunge transition
\cite{Ori:2000zn,Compere:2021iwh},
ii) in the ringdown phase it provides the rationale behind
the hidden symmetries (Darboux covariance,
Virasoro~\cite{Chandrasekhar:579245,Glampedakis:2017rar,Lenzi:2021njy,Lenzi:2021wpc,Lenzi:2022wjv,Lenzi:2023inn,Jaramillo:2024qjz}) related to the Korteweg-de Vries (KdV) equation
$\partial_t u =  6 u\partial_xu - \partial^3_xu$ underlying BH QNM-isospectral deformations,
and
iii) in the merger phase the (Hasting–Leod) Painlev\'e-II transcendent\footnote{Painlev\'e-II
is a `self-similar' solution of the (modified-)KdV, realising the Painlev\'e test
for KdV  integrability via inverse-scattering-transform, upon reduction to a linear problem.
In spite of KdV non-linearity, an essential linearity enters via
the `integrability-linearity  tie':
{\em ``Certain nonlinear problems have a surprisingly simple underlying structure, and
  can be solved by essentially linear methods''} \cite{AblSeg81}.
 Could such tie play a role in explaining `simplicity and universality' in BBH dynamics?}
provides the non-linear `turning point' scheme~\cite{AblSeg77} generalising the linear-Airy
one~\cite{wasow2018asymptotic}.

\medskip

{\bf Proposal of  Mechanism 2}. {\em Integrability of a (background) sector of the theory, with
Painlevé-II as a structural thread all along BBH dynamics: from inspiral to ringdown,
passing through the merger.}


\subsection{A `Wave-Mean Flow' approach and beyond: a hierarchical asymptotic  BBH dynamics program}
Integrability is not proposed above for the whole theory, but only in a background
sector of `slow' degrees of freedom (DoF). Linear and
non-linear elements are `reconciled' in a
``wave-mean flow'' approach, ultimately justified in a `soliton resolution'
picture \cite{tao2009solitons,bizon2022characteristic}: 
`fast' DoF ($\Psi$) linearly
propagating/interacting on a `slow'-DoF ($u$) integrable background.
In a sketchy manner, the dynamics splits as (cf. \cite{Jaramillo:2022oqn,Jaramillo:2023day,Jaramillo:2024qjz,Vitel25})
\bea
  \label{e:fast-slow_DoF}
  \left(-\Box + V(t,x;u) \right) \Psi =
  S(t,x;u) \qquad , \qquad
\partial_t u = F(t,x; u, u_x, u_{xx}, \ldots) \ .
\eea
These elements build up a hierarchical BBH
program~\cite{Jaramillo:2023day,Lenzi:2021njy,Lenzi:2021wpc,Jaramillo:2022mkh,Jaramillo:2022oqn,Lenzi:2022wjv,Lenzi:2023inn,Jaramillo:2024qjz,JarLam24}
(cf. table below).
\begin{center}
  {
    \scriptsize
    \begin{tabular}{|c||c|c|}
    \hline
    Asymptotic BBH Model & Mathematical/Physical Framework & Key Structures/Mechanisms \\
    \hline
    \hline
Fold-caustic model  & Geometric Optics                           & Arnol'd-Thom's Theorem  \\ 
                                & Catastrophe (singularity) Theory    & Classification of Stable Caustics   \\
    \hline
Airy function model  & Fresnel's Diffraction & Universal Diffraction Patterns \\
                                & Semiclassical Theory                      &    in Caustics \\
                                & Asymptotic ODE theory                 &  linear ODE turning points \\
\hline
Painlev\'e-II model    &  Painlev\'e Transcendents   & Painlev\'e property \\
& and Integrability &\\
                                &  Self-force calculations and EMRBs &  Non-linear Turning Points \\
\hline  
KdV-like model  &    Inverse Scattering Transform   &   Painlev\'e test, Lax pairs
\\
(Wave-Mean Flow)   &  and Integrability & Darboux transformations \\
&    Dispersive Non-linear PDEs                       &
 Scattering on Solitons, Soliton Resolution \\
&    Critical Phenomena    &  Universal Wave Patterns\\
& in Dispersive PDEs &  Dubrovin's Conjecture \\
\hline
Propagation models on  &  Ward's Conjecture  & (anti-)Self-Dual DoF \\
(anti)-Self-Dual & and Integrability  & Scattering on Instantons, Tunneling \\
 backgrounds  &       Twistorial techniques    &   Penrose Transform, `Twistor' BBH data   \\
\hline
    \end{tabular}
     }
\end{center}
Two complementary mechanisms have been proposed for simplicity and
universality in BBH dynamics. Then both answers may actually hold in our `boring versus elegant' question:
i) ``nothing special happens'' (boring) with Airy as a universal `caustic diffraction' 
(fast DoF, Mechanism 1), and 
ii) ``something special happens'' (elegant) with `Painlevé-II integrability' threading 
BBH dynamics (slow DoF, Mechanism 2).


\section*{Acknowledgments}
JLJ would like to thank the organisers of the 24th International
Conference on General Relativity and Gravitation (GR24) in Glasgow (14-18 July 2025), for the
opportunity to present this work. JLJ would also like to
thank Piotr Bizo\'n, Gregorio Carullo, J\"org Frauendiener, Shahar Hadar,
Abraham Harte, Vincent Lam,  Karim Mosani and Judy Shir
for discussions concerning the presentation at the GR24 conference. We acknowledge support from the
ANR ``Quantum Fields interacting with Geometry'' (QFG) project (ANR-20-CE40-0018-02),
from the EIPHI Graduate School (contract ANR-17-EURE-0002), from the
``Investissements d’Avenir'' program through project ISITE-BFC (ANR-15-IDEX-03) and
from the Spanish FIS2017-86497-C2-1 project (with FEDER contribution).

\bibliography{JARAMILLO_GR24_biblio}

\end{document}